\documentclass[12pt,epsf]{article}
\usepackage{graphicx}
\usepackage{epsfig}
\begin{document}
%%%%%%%%%%%%%%%%%%%%%%%%%%%%%%%%%%%%%%%%%%%

\def\a{\alpha}
\def\b{\beta}
\def\c{\varepsilon}
\def\d{\delta}
\def\e{\epsilon}
\def\f{\phi}
\def\g{\gamma}
\def\h{\theta}
\def\k{\kappa}
\def\l{\lambda}
\def\m{\mu}
\def\n{\nu}
\def\p{\psi}
\def\q{\partial}
\def\r{\rho}
\def\s{\sigma}
\def\t{\tau}
\def\u{\upsilon}
\def\v{\varphi}
\def\w{\omega}
\def\x{\xi}
\def\y{\eta}
\def\z{\zeta}
\def\D{\Delta}
\def\G{\Gamma}
\def\H{\Theta}
\def\L{\Lambda}
\def\F{\Phi}
\def\P{\Psi}
\def\S{\Sigma}

\def\o{\over}
\def\beq{\begin{eqnarray}}
\def\eeq{\end{eqnarray}}
\newcommand{\gsim}{ \mathop{}_{\textstyle \sim}^{\textstyle >} }
\newcommand{\lsim}{ \mathop{}_{\textstyle \sim}^{\textstyle <} }
\newcommand{\vev}[1]{ \left\langle {#1} \right\rangle }
\newcommand{\bra}[1]{ \langle {#1} | }
\newcommand{\ket}[1]{ | {#1} \rangle }
\newcommand{\EV}{ {\rm eV} }
\newcommand{\KEV}{ {\rm keV} }
\newcommand{\MEV}{ {\rm MeV} }
\newcommand{\GEV}{ {\rm GeV} }
\newcommand{\TEV}{ {\rm TeV} }
\def\diag{\mathop{\rm diag}\nolimits}
\def\Spin{\mathop{\rm Spin}}
\def\SO{\mathop{\rm SO}}
\def\O{\mathop{\rm O}}
\def\SU{\mathop{\rm SU}}
\def\U{\mathop{\rm U}}
\def\Sp{\mathop{\rm Sp}}
\def\SL{\mathop{\rm SL}}
\def\tr{\mathop{\rm tr}}

\def\IJMP{Int.~J.~Mod.~Phys. }
\def\MPL{Mod.~Phys.~Lett. }
\def\NP{Nucl.~Phys. }
\def\PL{Phys.~Lett. }
\def\PR{Phys.~Rev. }
\def\PRL{Phys.~Rev.~Lett. }
\def\PTP{Prog.~Theor.~Phys. }
\def\ZP{Z.~Phys. }

%%%%%%%%%%%%%%%%%%%%%%%%%%%%%%%%%%%%%%%%%%%%%%%%%%%%%%%%%%%%%%%%%%%%

\baselineskip 0.7cm

\begin{titlepage}

\begin{flushright}
UT-08-33\\
IPMU-08-0115
\end{flushright}

\vskip 1.35cm
\begin{center}
{\large \bf
   Cosmic Ray Positron and Electron Excess from Hidden-Fermion Dark Matter Decays
}
\vskip 1.2cm
Koichi Hamaguchi$^{1,2}$, Satoshi Shirai$^1$ and T.~T.~Yanagida$^{1,2}$
\vskip 0.4cm

{\it $^1$  Department of Physics, University of Tokyo,\\
     Tokyo 113-0033, Japan\\
$^2$ Institute for the Physics and Mathematics of the Universe, 
University of Tokyo,\\ Chiba 277-8568, Japan}

\vskip 1.5cm

\abstract{
The anomalies observed in recent cosmic ray experiments seem to strongly constrain the nature of the dark matter.
In this letter, we investigate a possibility of the fermionic dark matter with a minimal extension of the standard model.
We found that the dark matter decays caused by the dimension six operators can naturally explain the anomalies.
}
\end{center}
\end{titlepage}

\setcounter{page}{2}

\section{Introduction}

The presence of dark matter (DM) has been established by numerous
observations, which requires physics beyond the standard model (SM).
The nature of the DM is now one of the most important issues not only 
in cosmology but also in particle physics. 

The minimal extension of the SM is to introduce one extra particle called $X$, that is, 
a candidate particle for the DM \cite{Davoudiasl:2004be}. 
We assume the DM $X$ to be a singlet of the SM gauge groups.
The $X$ particle can be a boson or a fermion. 
We consider in this letter the case of fermion. We denote the DM $X$ as $\Psi$ to distinguish it 
from the case of boson \footnote{ See Refs. \cite{CTY} for the case of $X$ being a hidden gauge boson $A'$.}.

The new particle $\Psi$ may have Yukawa couplings as 
\beq
\lambda_i \Psi \ell^i H + {\rm h.c.} , \label{eq:2body}
\eeq
which induces too fast decay of the
$\Psi$. Here, $\ell$ is the lepton doublet, $H$ the Higgs doublet and
$i=1,2,3$ denotes family indices of the leptons. 
Thus, we assume the Yukawa coupling is strongly suppressed, otherwise the $\Psi$ can not be
a candidate of the DM. The required suppression may be easily obtained by suitable configurations of 
particle wavefunctions in a higher dimensional theory \footnote{
An example is given in a 5-dimensional space-time with an extra dimension $S^1/Z_2$. We put the $\Psi$
on one boundary and the Higgs $H$ on the other boundary. We put quarks and leptons in the bulk. Then,
we find the Yukawa coupling  $\Psi \ell H$ is exponentially suppressed 
if the size of the extra dimension
is large enough, compared with the inverse of the cut-off scale in the 5-dimensional theory.}.

If the above dangerous Yukawa coupling constants $\lambda_i$ in Eq.~(\ref{eq:2body}) 
are strongly suppressed as $|\lambda_i|\ll 10^{-26}$,
 the dominant operators become dimension six 
four-Fermi interactions among the SM particles and $\Psi$:
\begin{equation}
\frac{1}{M^2_*}\Psi \ell^i\ell^j {\bar e}^k, ~~~~\frac{1}{M^2_*}\Psi {\bar d}^i{\bar d}^j{\bar u}^k,
\end{equation}
where $i, j, k =1,2,3$ and $M_*$ is the cut-off scale, which would be regarded as the Grand Unification scale. The lifetime of the $\Psi$ is given by
\begin{equation}
\tau_\Psi \simeq 10^{26}~ {\rm sec} ~\left(\frac{M_*}{10^{15}~{\rm GeV}}\right)^4~\left(\frac{1~{\rm TeV}}{m_{\Psi}}\right)^5, \label{eq:life}
\end{equation}
where $m_{\Psi}$ is the mass of the DM $\Psi$. We see that the lifetime is much longer than the age of the universe
in a large parameter space of $M_*$ and $m_{\Psi}$ and hence the $\Psi$ can be a candidate of the DM.

The $\Psi$ may be produced non-thermally
in the early universe through the above dimension six interactions. 
The density parameter
of the $\Psi$ is given by
\begin{equation}
\Omega_{\psi} h^2 = {\cal O}(0.1) \left(\frac{T_R}{10^{11} {\rm GeV}}\right)^3~
\left(\frac{10^{15}~{\rm GeV}}{M_*}\right)^4~
~\left(\frac{m_\Psi}{1~{\rm TeV}}\right), \label{eq:density}
\end{equation}
where $T_R$ is the reheating temperature after the inflation \footnote{
The derivation of Eq.~(\ref{eq:density}) is as follows: 
The production cross section of the $\Psi$ is $\left<\sigma v\right>\approx T^2 M_*^{-4}$.
Following the Boltzmann equation, one can get $n_{\psi}/n_{\rm rad} \approx \left. n_{\rm rad} \left<\sigma v\right> H^{-1}\right|_{T=T_R}$. 
}.
We see that the observed DM density $\Omega_{\rm DM}h^2\simeq 0.1$  is also explained for a wide region of the parameter space of $M_*, m_\Psi$ and $T_R$.

The purpose of this letter is to show that the anomalous excess of cosmic ray electron and positron
recently observed by PAMELA~\cite{Adriani:2008zr} and ATIC~\cite{:2008zz}/PPB-BETS~\cite{Torii:2008xu} is naturally explained by the decay 
of the DM $\Psi$~\footnote{For recent progress in the study of the decaying DM signal, see Refs.~\cite{CTY,DMdecay-gravitino,DMdecay-composite,DMdecay-others}.}.
In the present analysis we assume the following dimension six operator, for simplicity;
\begin{equation}
 \frac{1}{M_*^2}[(\Psi {\bar e}^1) (\ell^1\ell^3)  + \alpha\;(\Psi {\bar u}^1) ({\bar d}^1{\bar d}^3)]+{\rm h.c}., \label{eq:3body}
\end{equation}
where $\alpha$ is a free parameter of ${\cal O}(1)$. We find that the replacement of the third family by
the second family does not significantly change our final result and hence we consider only the above operators.
However, the more general analysis including all possible dimension six operators is straightforward
and will be given elsewhere.

For completeness, we also study the case that the two-body decays of the DM are induced by Eq.~(\ref{eq:2body}).
We find that the required lifetime for the DM $\Psi$ is given 
if the Yukawa couplings $|\lambda_i|$ are sufficiently small as ${\cal O}(10^{-26})$.
We choose the flavor structure as $|\lambda_1|\gg |\lambda_{2,3}|$ in the present analysis.
\section{Cosmic rays from the hidden fermion DM decays}
Let us discuss the cosmic ray signals from the decays of the DM $\Psi$.
The interaction (\ref{eq:2body}) dominantly causes two-body decays of the DM $\Psi$,
\beq
\Psi \rightarrow h \nu,~Z \nu, ~ W^{\pm}e^{\mp} \label{eq:2}
\eeq
with branching ratio $1:1:2$.

The interaction (\ref{eq:3body}) mainly causes three-body decays
\beq
\Psi \rightarrow \tau^{\pm} e^{\mp} \nu,~e^{\pm} e^{\mp}\nu,~dbu,~\bar{d}\,\bar{b}\,\bar{u} \label{eq:3}
\eeq
with branching ratio $1:1:3\alpha^2:3\alpha^2$.
Here, we have assumed that the $\Psi$ is a Majorana fermion and that  $m_h \simeq 110$ GeV and $m_{\Psi}\gg m_{h}$.
In both cases Eqs.~(\ref{eq:2}) and (\ref{eq:3}), high energy electron, positron, photon and antiproton are produced.
To estimate the energy spectrum of these decay products, we have used the program PYTHIA~\cite{Sjostrand:2006za}. 
The particles produced in the DM decays are influenced by various factors in the propagation.
For the propagation in the Galaxy, we adopt the method discussed in Refs.~\cite{Ibarra,Hisano:2005ec} with
Navarro, Frenk and White halo profile~\cite{Navarro:1996gj};
\beq
\rho_{DM}=\frac{\rho_0}{(r/r_c)[1+(r/r_c)]^2},\label{eq:NFW}
\eeq
where $\rho_0=0.26~ {\rm GeV cm^{-3}}$ and $r_c= 20~{\rm kpc}$.

\subsection*{Positrons and electrons}
\begin{figure}[h!]
\begin{tabular}{cc}
\begin{minipage}{0.5\hsize}
\begin{center}
\epsfig{file=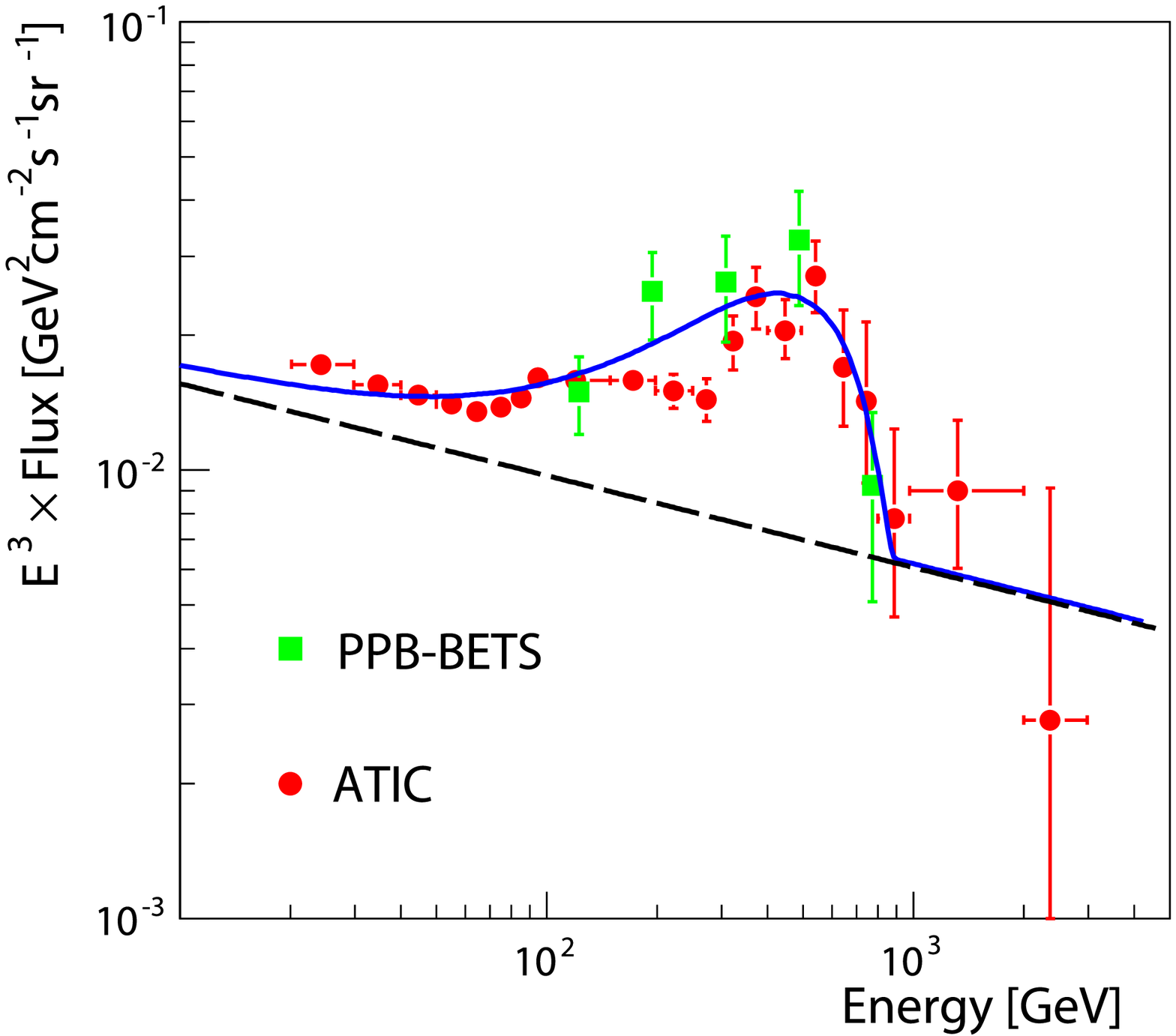 ,scale=.42,clip}
\end{center}
\end{minipage}
\begin{minipage}{0.5\hsize}
\begin{center}
\epsfig{file=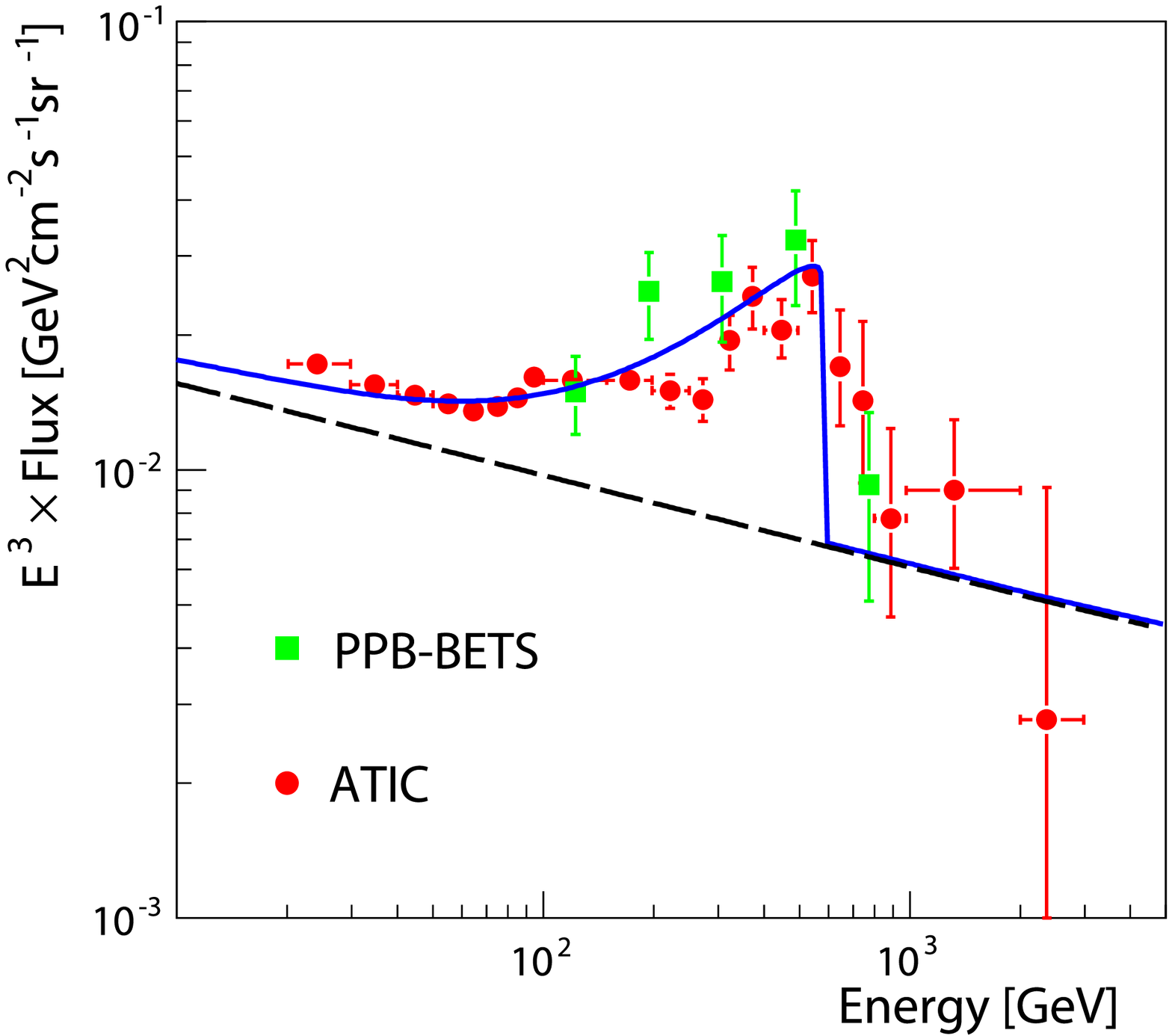 ,scale=.42,clip}
\end{center}
\end{minipage}
\end{tabular}
\caption{Positron and electron fluxes with experimental data~\cite{:2008zz,Torii:2008xu}.
 Left: three-body decay. Right: two-body decay.
The solid line represents the DM signal plus background and the dashed line the background.
}
\label{fig:FLUX}
\end{figure}

As a diffusion model, we use MED model in Ref.~\cite{Delahaye:2007fr}.
We see that the positrons and electrons come from the DM decays inside the Galaxy, 
and especially the decays within a few kpc 
from us are important.
In Fig.~\ref{fig:FLUX}, we show the total flux of the electron and positron.
The left figure shows the case that the interaction (\ref{eq:3body}) is a dominant interaction causing three-body decay.
We set the DM mass $m_{\Psi}=1800$ GeV, the lifetime $\tau_{\Psi}=9\times10^{25}$ sec and $\alpha=(2\sqrt{2})^{-1}$.
Hereafter we fix $\alpha=(2\sqrt{2})^{-1}$.
The lifetime is given by $M_* \simeq 3\times 10^{15}$ GeV in Eq.~(\ref{eq:life}).
For the right figure, the case that the interaction (\ref{eq:2body}) is dominant is shown.
Here, we set the DM mass $m_{\Psi}=1200$ GeV, the lifetime $\tau_{\Psi}=8\times10^{25}$ sec.
The lifetime is obtained by $\lambda_1 \simeq 1\times 10^{-26}$ in Eq.~(\ref{eq:2body}).
As for the background flux, we set $0.0253(E/1~\GEV)^{-3.206}~{\rm GeV^{-1} cm^{-2}s^{-1}sr^{-1}}$ in both cases \footnote{
This background is estimated by fitting the data points as BG + signal, assuming BG is power-low and adopting the signal in the 
case of three-body decay.
As the fitting parameters, we have used the lifetime of the $\Psi$, $\alpha$ in Eq. (\ref{eq:3body}), and the coefficient and power of the background.
Here, we set the weight for each data point $1$.
}
.
\begin{figure}[h!]
\begin{tabular}{cc}
\begin{minipage}{0.5\hsize}
\begin{center}
\epsfig{file=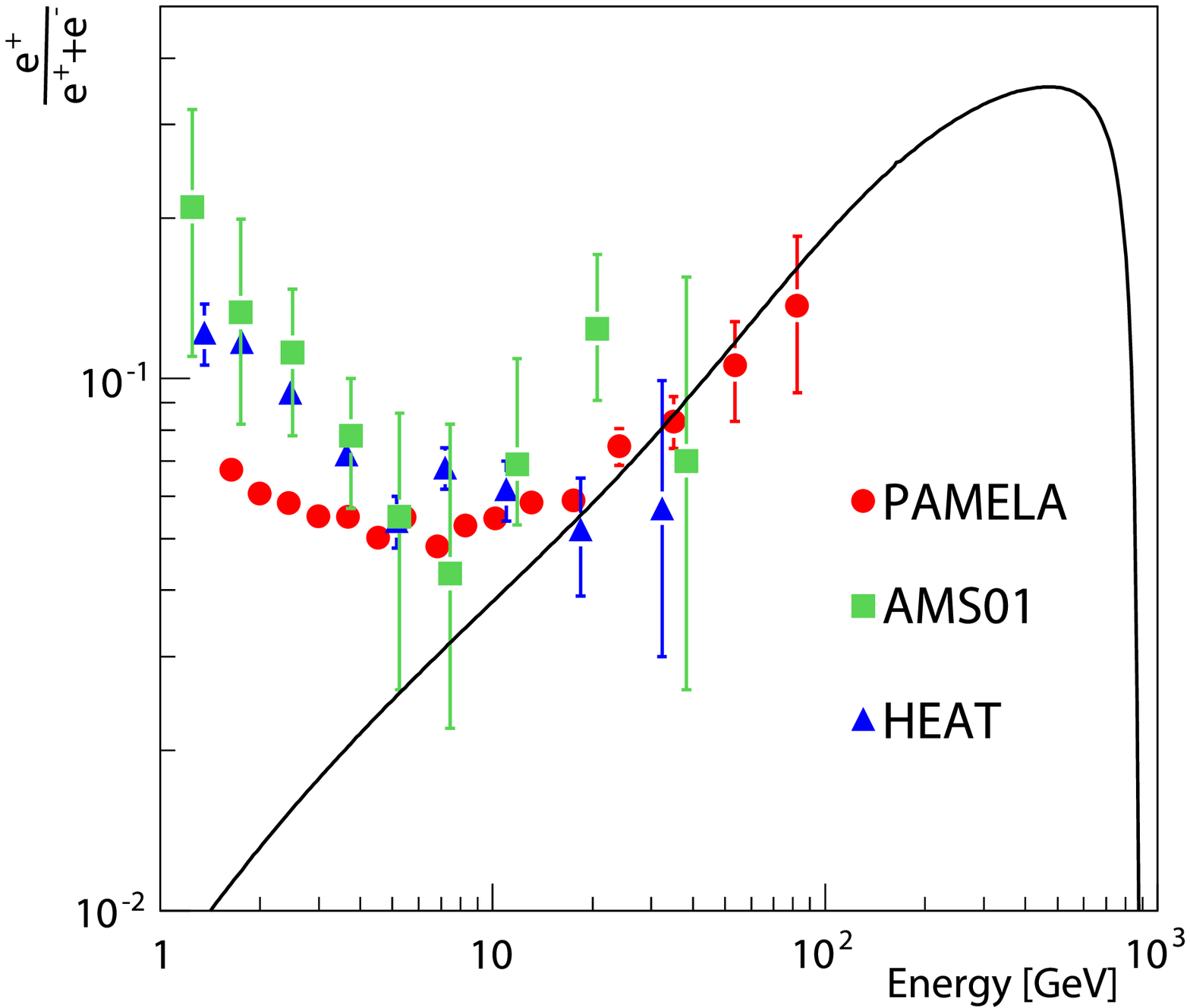 ,scale=.42,clip}
\end{center}
\end{minipage}
\begin{minipage}{0.5\hsize}
\begin{center}
\epsfig{file=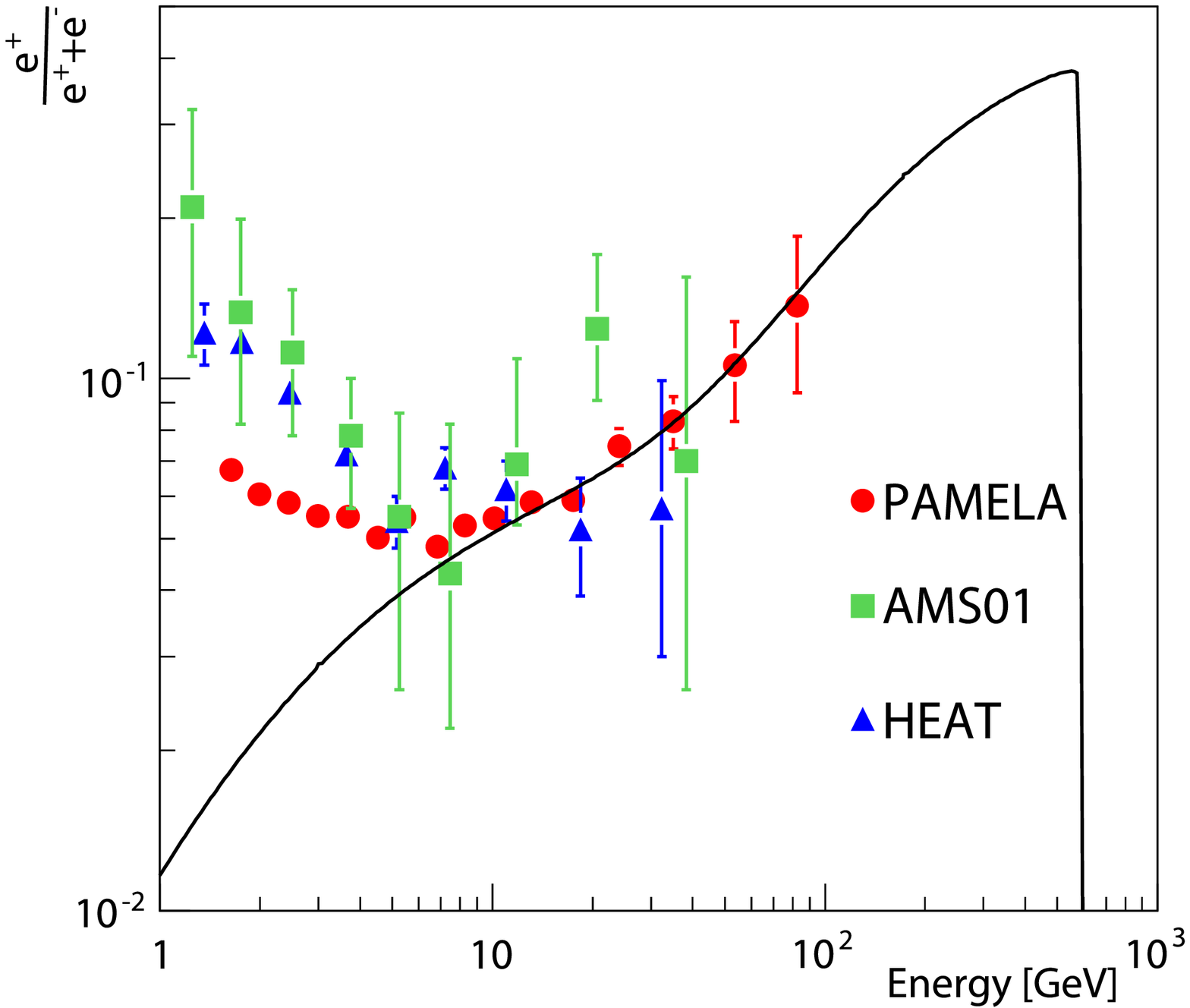 ,scale=.42,clip}
\end{center}
\end{minipage}
\end{tabular}
\caption{Positron fractions with experimental data \cite{Adriani:2008zr,Aguilar:2007yf,Barwick:1997ig}.
 Left: three-body decay. Right: two-body decay.}
\label{fig:FRAC}
\end{figure}

Next, we estimate the positron fraction.
For the background, we extrapolate the background estimated in Fig.~\ref{fig:FLUX} and assume that the background
consists entirely of the electron, since the secondary positron would be negligible for $E\gsim 10$ GeV. 
Fig.~\ref{fig:FRAC} shows the positron fraction.
One can see good agreements with the experimental data in both cases except in a low-energy region.
The behavior of the background in the low-energy region is complicated due to various factors such as solar modulation or contamination from the
secondary positron.
This can be reasons why the naive background estimation is not good in the low energy region.
The detailed treatment of the background is out of the reach of this letter.

\subsection*{Gamma ray}
For the gamma ray, both of the DM decays in the halo and extra-Galaxy are important.
To estimate the halo component, we have used the NFW profile in Eq.~(\ref{eq:NFW}) 
and averaged the halo signal over the whole sky excluding the region within $\pm 10^\circ$ around the Galactic plane.

For the extragalactic component, the gamma ray is influenced by the red-shift.
We estimate the extragalactic component by using the following cosmological parameters;
$\Omega_{\Psi}h^2\simeq0.11,~\Omega_{\rm matter}h^2\simeq0.13,~\Omega_{\Lambda}\simeq0.74,~\rho_c \simeq 1.0537\times 10^{-5} h^2~ {\rm GeV cm^{-3}},~h\simeq 0.72$~\cite{Komatsu:2008hk}.

\begin{figure}[htbp]
\begin{tabular}{cc}
\begin{minipage}{0.5\hsize}
\begin{center}
\epsfig{file=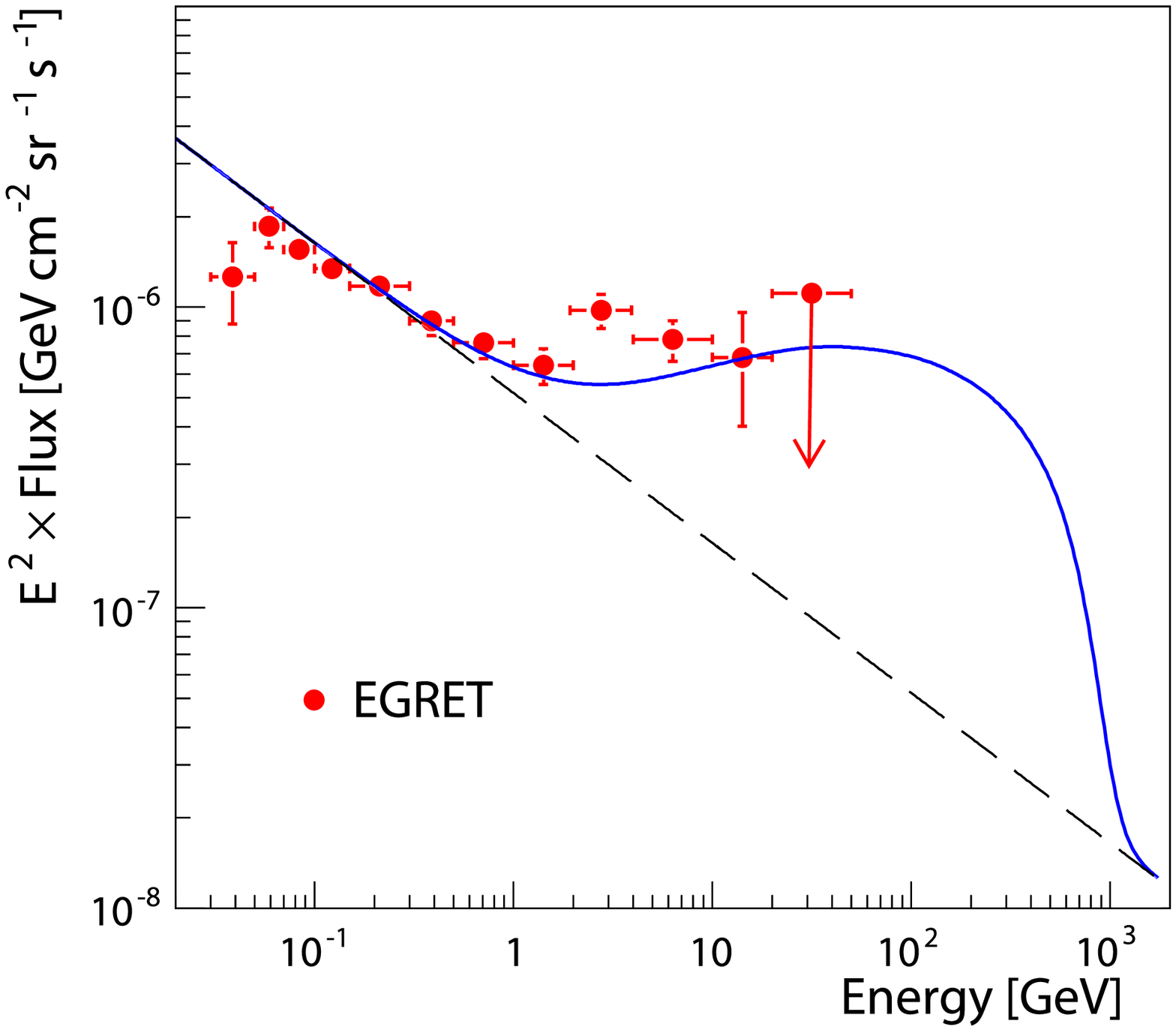 ,scale=.42,clip}
\end{center}
\end{minipage}
\begin{minipage}{0.5\hsize}
\begin{center}
\epsfig{file=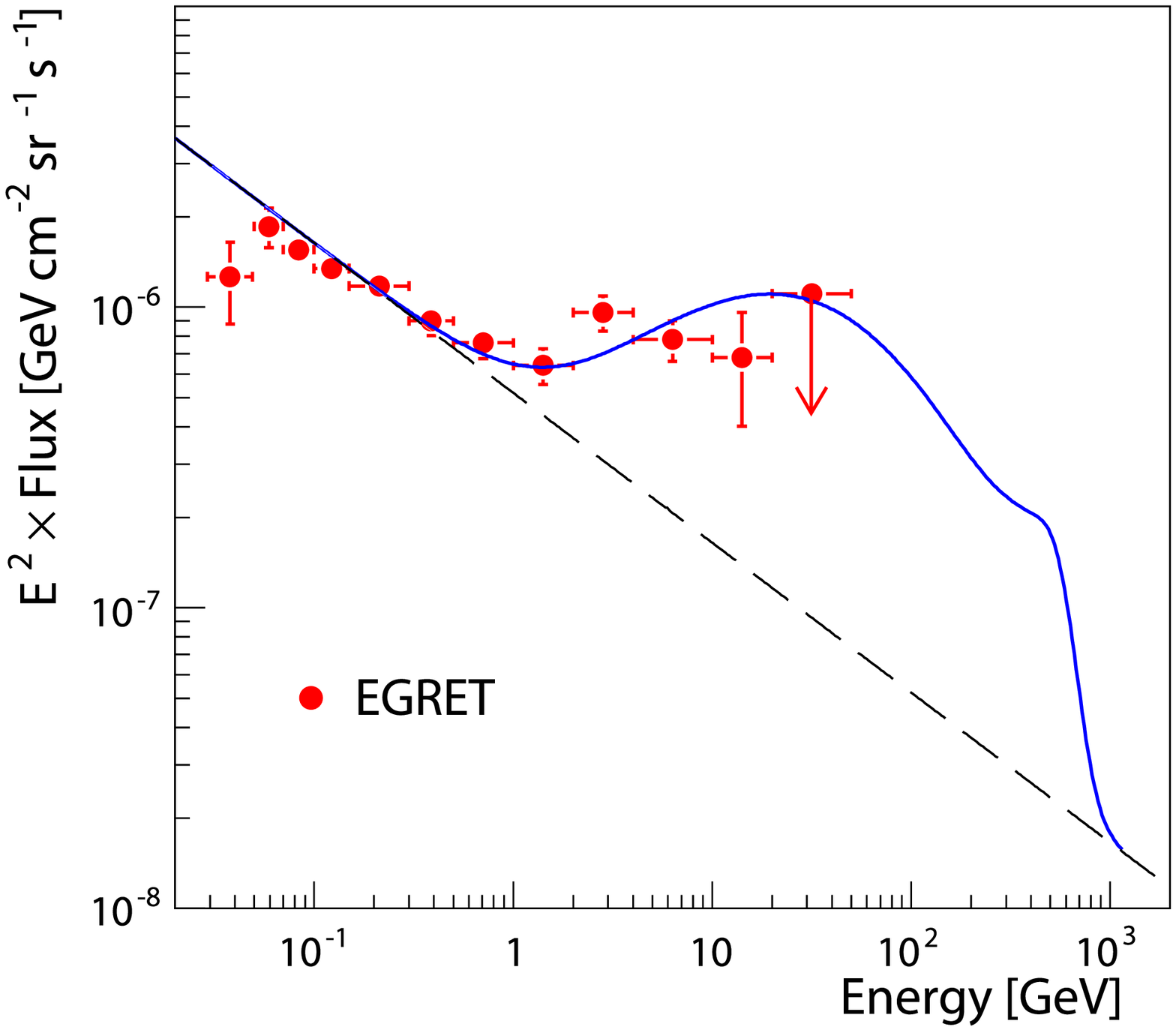 ,scale=.42,clip}
\end{center}
\end{minipage}
\end{tabular}
\caption{Gamma ray fluxes with experimental data~\cite{Sreekumar:1997un,Strong:2004ry}. Left: three-body decay. Right: two-body decay.
Solid line represents the DM signal plus background and dashed the background only.}
\label{fig:GFLUX}
\end{figure}

In Fig.~\ref{fig:GFLUX}, the gamma ray fluxes are shown.
We set the background flux as $5.18\times 10^{-7}(E/1~\GEV)^{-2.499}~{\rm GeV^{-1} cm^{-2}s^{-1}sr^{-1}}$ as in Ref.~\cite{Ishiwata:2008cu}.
We have assumed the energy resolution is $15 \%$.
In both cases, the DM signals are consistent with the current experiment data and 
anomalous behavior of the gamma ray flux is expected to continue up to higher energy.
This will be tested by upcoming Fermi Gamma-Ray Space Telescope (FGST, formerly named GLAST~\cite{FGST}).

\subsection*{Antiproton}
\begin{figure}[htbp]
\begin{tabular}{cc}
\begin{minipage}{0.5\hsize}
\begin{center}
\epsfig{file=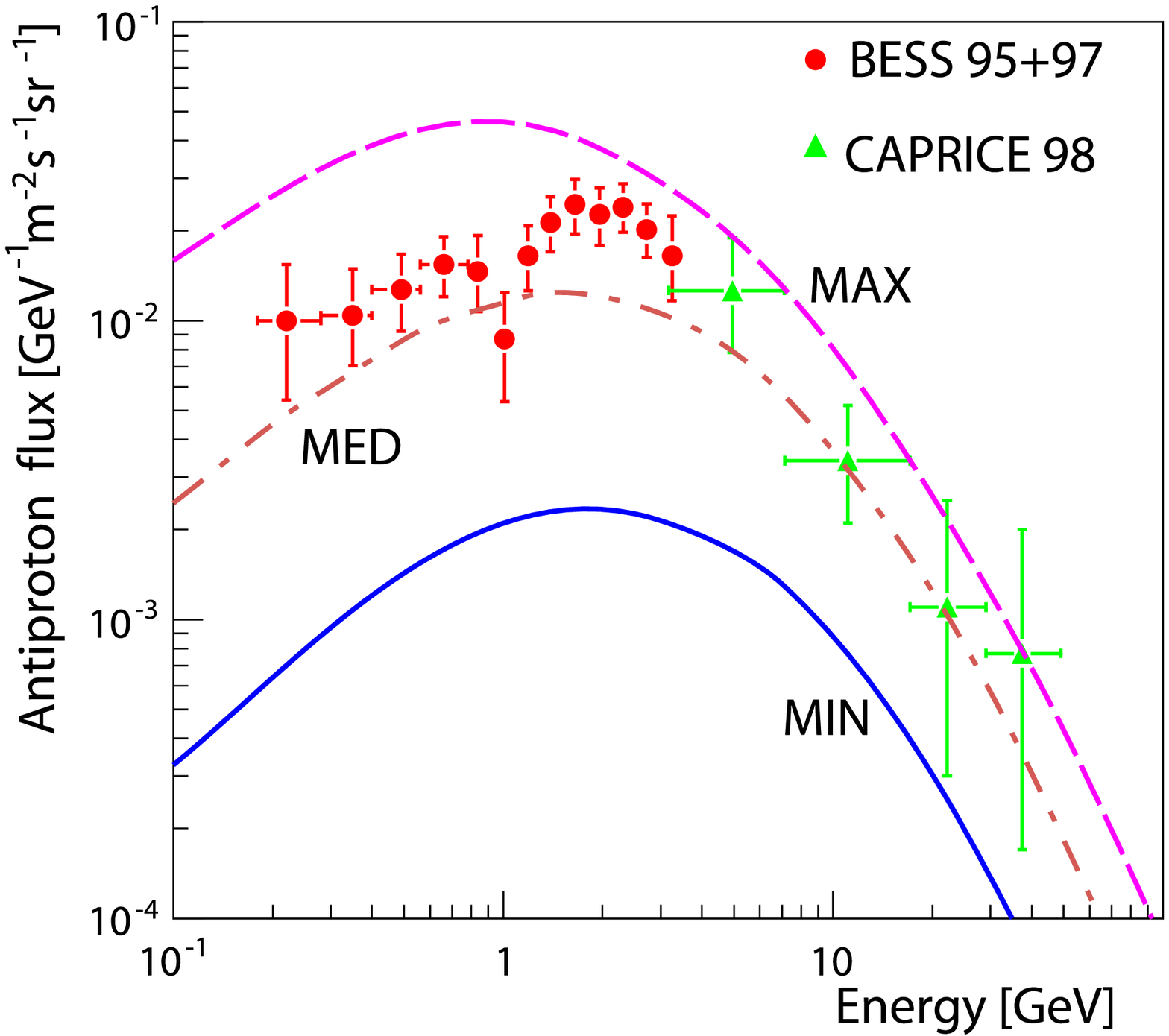 ,scale=.42,clip}
\end{center}
\end{minipage}
\begin{minipage}{0.5\hsize}
\begin{center}
\epsfig{file=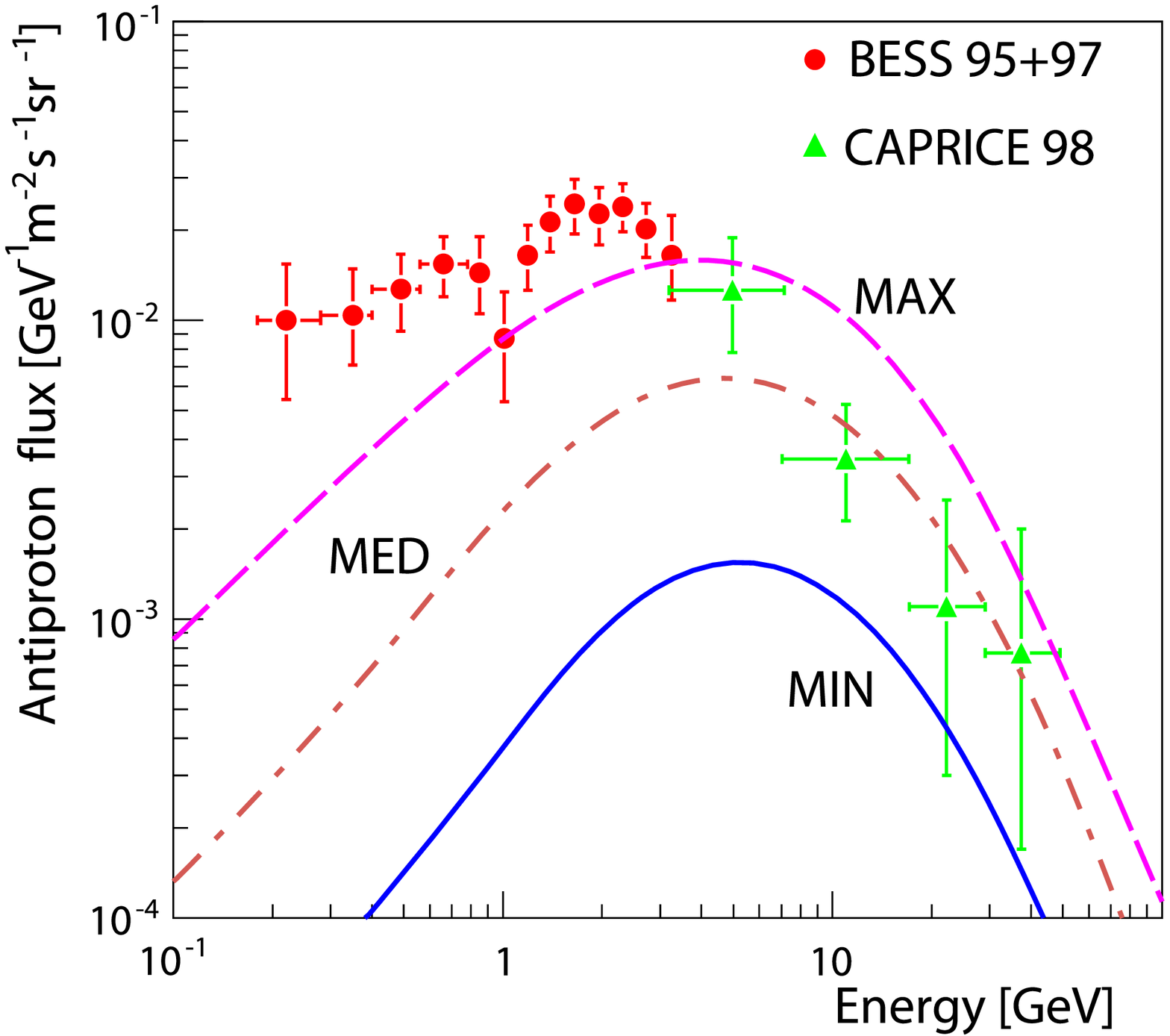 ,scale=.42,clip}
\end{center}
\end{minipage}
\end{tabular}
\caption{Antiproton fluxes with experimental data~\cite{Matsunaga:1998he,Boezio:2001ac}. Left: three-body decay.  Right: two-body decay.
Solid, dash-dotted and dashed line represent MIN, MED and MAX diffusion models, respectively. }
\label{fig:ANTIFLUX}
\end{figure}

We estimate the antiproton flux, following the Refs.~\cite{Ibarra}.
For the solar modulation, we set $\phi_F=500$ MV.
In Fig.~\ref{fig:ANTIFLUX}, we show the antiproton fluxes for some different diffusion models~\cite{Delahaye:2007fr}.
Here, we show only the DM signals.
We can see the contradiction between the experiments and the signals in some diffusion models.
However, in both cases of the two- and three-body decay, MIN models (and also MED model for the three-body decay case) do not
conflict with the experimental data.
Therefore, for the antiproton, the DM $\psi$ is consistent with the experimental data at least in some diffusion models.

\section{Discussion and conclusions}
In this letter, we have investigated the case of the fermionic DM.
Both two- and three-body decays can explain the anomaly of electron and positron cosmic ray.
In addition, they are also consistent with the gamma ray anomaly.
Especially, the decay caused by the dimension six operators seems to be attractive, since
it naturally explains the proper lifetime of the DM for the GUT-scale cut-off $M_*\simeq 10^{15}$--$10^{16}$ GeV\footnote{DM decays via GUT-scale physics have been also discussed in a recent work~\cite{Arvanitaki:2008hq,Nardi:2008ix}.}.

There are remaining issues.
For example, the reason of large suppression of Eq.~(\ref{eq:2body}) is unclear.
In addition, the reason why the DM $\Psi$ decays dominantly into the first family leptons is unclear.
However this problem may be solved by choosing suitable Froggatt-Nielsen charge for the DM $\Psi$
\footnote{For example, we can consider a discrete $Z_6$ symmetry~\cite{Fujii:2001zr}. Since the $\Psi$ is a Majorana particle, 
the $\Psi$ should have $3$ of the $Z_6$ charge. 
The $Z_6$ charges of other particles are 
$(\ell_1,\ell_2,\ell_3);(1,0,0)$,
$(\bar{e}_1,\bar{e}_2,\bar{e}_3);(2,1,0)$, 
$(q_1,q_2,q_3);(2,1,0)$, 
$(\bar{u}_1,\bar{u}_2,\bar{u}_3);(2,1,0)$, 
$(\bar{d}_1,\bar{d}_2,\bar{d}_3);(1,0,0)$, $H;0$.
This choice of Froggatt-Nielsen charge leads to the 
$\Psi$ decays considered in the text.
}.

Finally, we should note that the hidden fermion $\Psi$ can be identified as the lightest 
neutralino in the supersymmetric standard model.
In this case, R-parity breaking operators may induce the DM $\Psi$ decays in similar ways discussed
in the present letter.
\paragraph{\it Note Added:} Discrimination between two- and three-body decay of the DM 
would be possible in the future cosmic ray experiments~\cite{Chen:2008fx}.
\section*{Acknowledgement}
KH thanks to C.~R.~Chen, M.~M.~Nojiri and F.~Takahashi for discussions.
This work was supported by World Premier International Center Initiative (WPI Program), MEXT, Japan. 
The work of SS is supported in part by JSPS Research Fellowships for Young Scientists. 

\end{document}